\documentclass[twocolumn]{emulateapj}

\newcommand{\Myr}{{\rm \, Myr}}
\newcommand{\pc}{{\rm \, pc}}
\newcommand{\kpc}{{\rm \, kpc}}
\newcommand{\cm}{{\rm \, cm}}
\newcommand{\s}{{\rm \, s}}

\newcommand{\pder}[2]{\frac{\partial {#1}}{\partial {#2}}}

\begin{document}

\title{Can the cosmic-ray driven dynamo model
explain the observations of the polarized emission of edge-on galaxies ?
}
\author{Katarzyna Otmianowska-Mazur\altaffilmark{1},
Marian Soida\altaffilmark{1},
Barbara Kulesza-\.Zydzik\altaffilmark{1},
Micha{\l} Hanasz\altaffilmark{3},
and Grzegorz Kowal\altaffilmark{2} }
\altaffiltext{1}{Astronomical Observatory, Jagiellonian University,\\
ul. Orla 171, 30-244 Krak\'{o}w, Poland}
\altaffiltext{2}{Department of Astronomy, University of Wisconsin,\\
475 North Charter Street, Madison, WI 53706, USA}
\altaffiltext{3}{Toru\'{n} Centre for Astronomy, Nicolaus Copernicus University,\\
ul. Gagarina 11, 87-100 Toru\'{n}, Poland}

\begin{abstract}

In the present paper we construct maps of polarized synchrotron radio emission of a whole galaxy, based on local models of the cosmic ray (CR) driven dynamo. We perform numerical simulations of the dynamo in local Cartesian domains, with shear-periodic boundary conditions, placed at the different galactocentric radii. Those local solutions are concatenated together to construct the synchrotron images of the whole galaxy. The main aim of the paper is to compare the model results with the observed radio continuum emission from nearly edge-on spiral galaxy.

On the basis of the modeled evolution of the magnetic field structure, the polarization maps can be calculated at different time-steps and at any orientation of the modeled galaxy. For the first time a self-consistent cosmic-ray electron distribution is used to integrate synchrotron emissivity along the line of sight. Finally, our maps are convolved with the given radiotelescope beam. We show that it is possible to reconstruct the  extended magnetic halo structures of the edge-on galaxies (so called X-shaped structures).

\end{abstract}


\section{Introduction}
\label{sec:intro}

Recent deep observations of polarized radio-continuum emission at centimeter
wavelengths of edge-on spiral galaxies \cite[]{tul00,soida05,heesen05} revealed
characteristic  X-shaped magnetic field structure. Many galaxies showed magnetic
field oriented in the halo at a large angle to the galaxy disk. Earlier
observations, mostly limited by sensitivity to narrow area along the major axis,
demonstrated mainly plane-parallel component \citep[e.g.][]{dumke95}. The
polarization vectors of  NGC~5775 close to the galactic plane are parallel to
the disk, but in the halo they extend up to 10~kpc above the disk forming so
called $X$-shaped structure in all four quadrants. Such picture most probably
reflects  the quadrupole configuration of magnetic field in NGC~5775
\cite{soida08}. The observed degree of polarization is very high (locally
reaching even up to 50\%). It resembles high regularity of magnetic field there.
Such structures seem to be common among edge-on spiral galaxies (NGC~5775,
NGC~4666, NGC~4217 NGC~3628, NGC~253, NGC~891, \citep[see
e.g.][]{dahlem97,sukumar91,soida05}. In addition, \cite{tul00} detected that
NGC~5775 shows differential rotation in the direction perpendicular to the disk.
So far, there is no physical explanation for those extended structures.

An attempt to explain the structure of polarized vectors by the dipolar dynamo
wave in the NGC~5775 was made by \cite{sokol02}. He used a solution of the
dipolar classical turbulent dynamo equation and as a result he got structures
not confirmed by the observations so far \cite{soida08}.

The principle of the action of the CR-driven dynamo
\citep{parker92,hanasz00,hanasz04} is based on the cosmic ray (CR) energy
supplied continuously by supernova (SN) remnants. Due to the anisotropic
diffusion of cosmic rays along the horizontal magnetic field lines, cosmic rays
tend to accumulate within the disc volume. However, the configuration stratified
by the vertical gravity is unstable against the Parker instability. Buoyancy
effects induce the vertical and horizontal motions of the fluid and the
formation of undulatory patterns -- magnetic loops in the frozen-in,
predominantly horizontal magnetic fields. The presence of rotation in galactic
disks implies a coherent twisting of the loops by means of the Coriolis force
leading to the generation of the small-scale radial magnetic field components.
The next phase is merging the small-scale loops by the magnetic reconnection
process which forms the large scale radial magnetic fields. Finally, the
differential rotation stretches the radial magnetic field amplifying the
large-scale azimuthal magnetic field. The coupling of amplification processes of
the radial and azimuthal magnetic field components results in exponential growth
of the large scale magnetic field with timescales of galactic rotation period
(140~Myr) as shown by \cite{hanasz06}.

In this paper we present the results of galactic disk modeling which is
constructed from a set of local volumes placed at different distances from the
galactic center. We show the results of integration of synchrotron emissivities
(Stokes parameters I, Q and U) along the line of sight (LOS), and compare them
to the observed radio images of the galaxy NGC~5775. Our results are compared to
polarized radio continuum observations made at 6\,cm \citep{tul00}.

%
\section{Models of the cosmic-ray driven dynamo}
\label{sec:numerical_model}


The first complete 3D numerical model of the CR-driven dynamo has been
demonstrated by \cite{hanasz04,hanasz05,hanasz06,otm07}. Our  model of the
cosmic ray driven dynamo includes the following physical elements: the ionized
gas and magnetic field described by resistive MHD equations, the cosmic ray
component described by the diffusion-advection transport equation \cite[see][for
the details of the numerical algorithm]{hanasz03}, cosmic rays diffusing
anisotropically along magnetic field lines \citep{giacalone99,jokipii99},
supernova remnants exploding randomly in the disk volume, the finite (currently
uniform) resistivity of the ISM \cite[see][]{hanasz02,hanasz03,kowal03,tanuma03}
and the realistic vertical disk gravity and rotation \citep{ferriere98}.
In the present models gas motions arise only from the cosmic-ray
pressure gradients. 
We neglect any shock effects, like gas heating. It is caused by limitations
of currently used explicit
algorithm for cosmic ray difusion, although we plan to incorporate shock effects
in our future papers. A complementary work incorporating affects of dynamo
powered directly by supernova-driven turbulence, but without taking into account
the cosmic rays, has been recently published  by \cite{gressel08a,gressel08b}. In
the present paper we attempt to examine observational properties of our model of
cosmic ray-driven dynamo.

The system of coordinates $x,y,z$ corresponds locally to the global galactic
cylindrical system $r, \phi, z$. 
The boundary conditions are periodic in the Y-direction, sheared in the
X-direction \cite[following][]{hawley95} and open in the Z-direction for fluid
quantities and magnetic field components. The boundary conditions for cosmic
rays are fixed ($e_{cr} = 0$) on the Z-boundaries.

We present a new set of numerical models of the CR-driven dynamo computed in
cubes situated at different distances from the center of the galaxy (see Fig.~1).
We intend to obtain the picture of the whole galaxy from
these local cubes. In the inner part of the galaxy a variety of different
physical mechanisms occur (such as central activity, bulge influence etc.),
most of which are not taken into account by our model.  For this reason, we
start our calculations at 2~kpc radius. The physical parameters used in the
calculations are assumed according to the paper of \cite{ferriere98} which
refers to the Milky Way (see Fig.~1).
Our numerical simulations are performed using the Zeus-3D MHD code
\citep{stone92a,stone92b} with  the cosmic rays  extension made by 
Hanasz $\&$ Lesch  (2003).

The full set of equations describing the model includes the set of resistive MHD
equations completed by the cosmic ray transport  equation \cite[see][]{hanasz04}

\begin{equation}\pder{\rho}{t}{}+ \nabla (\rho V) = 0, \label{eqofconti}
\end{equation}
\begin{equation}
\pder{e}{t}{} +\nabla\cdot \left( e V\right)
    = - p \left( \nabla \cdot V
\right), \label{enereq}
\end{equation}
\begin{eqnarray}
\pder{V}{t}{} + (V \cdot \nabla)V
   = -\frac{1}{\rho}{} \nabla
   \left(p + p_{\rm cr} + \frac{B^2}{8\pi}\right)   \nonumber \\
   + \frac{B \cdot \nabla B}{4 \pi \rho}{}
      -2 \Omega \times V
   + 2 q {\Omega}^2 x \hat{e}_x + g_z(z)\hat{e}_z,
   \label{eqofmot}
\end{eqnarray}

\begin{equation}
\pder{B}{t}{} = \nabla \times \left( V \times B\right)
+ \eta \Delta B
\label{indeq},
\end{equation}

\begin{equation}
p=(\gamma-1) e, \quad \gamma=5/3
\end{equation}
where $\rm q=- \rm{d~ln\Omega/d~lnR}$ is the shearing parameter,  ($\rm R$ is
the distance to galactic center), $\eta $ is the resistivity, $\gamma$ is the
adiabatic index of thermal gas,  the gradient of cosmic ray pressure  $\nabla
p_{cr}$ is included in the equation of motion \cite[see][e.g.]{berezinski90} and
other symbols have their usual meaning. The uniform resistivity is included only
in the induction equation \cite[see][]{hanasz02}. The thermal, ionized gas
component is treated as an adiabatic medium.

The transport of the cosmic ray component is described by the diffusion-advection equation
\begin{equation}
\pder{e_{\rm cr}}{t}{} +{\nabla }\left( e_{\rm cr} {V}\right)
= {\nabla} \left(\hat{K} {\nabla} e_{cr} \right)
- p_{\rm cr} \left( {\nabla} \cdot {V} \right)
+ Q_{\rm SN}, \label{diff-adv-eq}
\end{equation}
where $Q_{\rm SN}$ represents the source term for the cosmic  ray energy density: the rate of production of cosmic rays injected  locally in SN remnants and
\begin{equation}
p_{\rm cr}=(\gamma_{\rm cr}-1) e_{\rm cr}, \quad \gamma_{\rm cr}=14/9.
\end{equation}
The adiabatic index of the cosmic ray gas $\gamma_{\rm cr}$ and the formula for
diffusion tensor
\begin{equation}
K_{ij} = K_{\rm \perp} \delta_{ij} + (K_\parallel - K_{\rm \perp}) n_i n_j,
\quad n_i = B_i/B,
\label{diftens}
\end{equation}
are adopted following the argumentation by \cite{ryu03}.

In order to construct the polarization maps of an edge-on galaxy characterized
by a high halo, our local simulation is extended to 4\,kpc above and below the
disk. The local cube X-size is  500\,pc, while Y-size is
1000\,pc. The cell sizes are equal to dx=dy=dz=20\,pc.
The resolution of our local calculations is given by 25 $\times$ 50 $\times$
400 grid points. However, we are
primarily interested is in the large scale magnetic field amplification and
structure. Because CR gas propagates diffusively and gas dynamics does not
involve shock waves, the currently assumed grid resolution is sufficient. This
statement is verified by resolution studies - simulations performed with the
cell size $(10 \pc)^3$ up to $(20 \pc)^3$ lead to convergent results.

\begin{figure}
 \begin{center}
  \includegraphics[height=0.40\textheight]{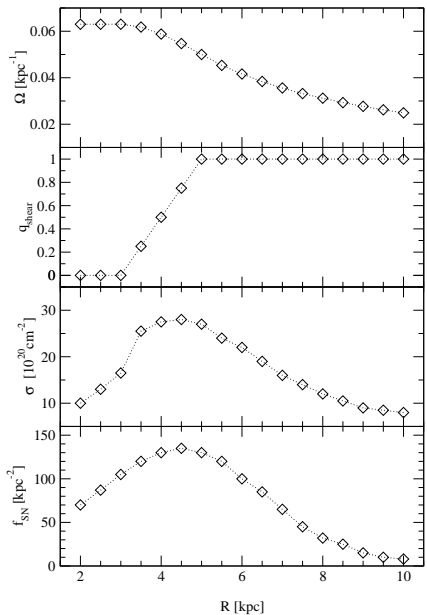}
 \end{center}
 \caption{
Radially dependent parameters for all our local calculations. Subsequent panels (counting
from the top) display angular velocity $\Omega$, shearing parameter $q_{shear}$,
gas column density $\sigma$ and supernova rate $f{_SN}$, vs. galactocentric
radius $R_G$, deduced from the model
by Ferriere (1998). 
}
\label{input}
\end{figure}

The disk rotation was defined by the values of
the angular velocity $\Omega$ at the different galactocentric radii ranging from
$0.05~\Myr^{-1}$ at $R_G=5 \kpc$ down to $0.025~\Myr^{-1}$ at $R_G=10 \kpc$
(see. Fig.~1). 
The value of the shearing parameter is $q=0$ up to a radius of 
3\,kpc, and linearly increasing to $q=1$ at 5\,kpc, as it is presented in
Fig.~1.
The cosmic ray diffusion coefficients assumed in the
simulations are: $K_\parallel=3 \times 10^{27} \cm^2\s^{-1}$ and $K_\perp =3
\times 10^{26} \cm^2\s^{-1}$. These values are scaled down by an order of
magnitude with respect to expected realistic values \citep[e.g.][]{jokipii99}
due to the timestep limitation in the currently used explicit algorithm for the
diffusion equation. The assumed value of the resistivity coefficient $\eta$ is
$3\times 10^{25} \cm^2\s^{-1}$. The values of other input parameters are
depicted in Fig.~\ref{input}.

\section{Construction of synthetic radio maps}

As the result of our local calculations we obtained 3D rectangular cubes of the
magnetic field and the cosmic-ray energy density. Since the CR component is
described as adiabatic relativistic gas instead of the full momentum
distribution function, the energy density of cosmic rays can be considered as
proportional to the number density of CR nucleons. The number density of cosmic
ray electrons $n_{e,cr}$ is typically assumed to be of the order of 1~\% of the
number density of cosmic ray nucleons $n_{n,cr}$. Therefore $n_{e,cr}$ can be
assumed proportional to $e_{cr}$. The local cubes at given radius and chosen
time-step are replicated into subsequent cylinders, and cylinders are combined
together into the full galactic disk (at the radii 2 -- 10\,kpc, and of 8\,kpc
thickness). The synthesized disk can be oriented according to the inclination
and position angle of any real galaxy (in the case of NGC\,5775 it is 86$^\circ$
and 145$^\circ$, respectively). After computing the magnetic field component
perpendicular to the line of sight $B_\perp$ and relativistic-electron density
$n_{e,cr}$ from the simulation data we calculate the synchrotron emissivity
at each point. Following the standard formula \citep[see e.g.][]{longair94} we
obtain Stokes parameters I, Q, and U:
\begin{equation}
{d \over d l}\left( \begin{array}[c]{c} I\\ Q\\ U \end{array} \right)
= \left( \begin{array}[c]{ccc}
\epsilon_I & 0 & 0\\
p \epsilon_I \cos 2\chi & \cos\Delta & -\sin\Delta\\
p \epsilon_I \sin 2\chi & \sin\Delta & \cos\Delta
\end{array}\right)
\left( \begin{array}[c]{c} 1\\ Q\\ U \end{array} \right),
\end{equation}
where the synchrotron emissivity is
\begin{equation}
 \epsilon_I\propto n_{e,cr}B_\perp^{(\gamma+1)/2} 
 \propto e_{cr}B_\perp^{(\gamma+1)/2},
\end{equation}
and $\Delta$ denotes the Faraday rotation angle.
Integrating the Stokes parameters along the line of sight we obtain the map of
synchrotron emission of the simulated galaxy. We set $\gamma=2.8$ and $p=75\%$.

In addition, assuming that the distribution of thermal electrons $n_{e,th}$ is
proportional to the gas density, we can account for the Faraday rotation of
polarized emission along the line of sight. The Faraday rotation in small
distance $d l$ is
\begin{equation}
 \Delta\propto n_{e,th} B_\| dl.
\end{equation}

Finally, we convolve resulting I, Q, and U maps with Gaussian beam of HPBW of
20\arcsec{} and calculate polarized intensity and vectors of magnetic
polarization ($B$) for direct comparison with real observations of NGC~\,5775.

\section{Results}

\begin{figure}
 \begin{center}
  \includegraphics[height=0.40\textheight]{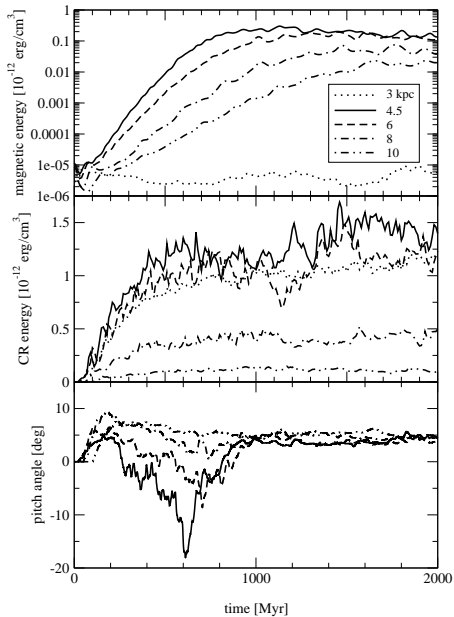}
 \end{center}
 \caption{Evolution of the magnetic field energy, the CR energy and the mean
pitch angle values at five chosen distances (R) from the galactic center: 3\,kpc
(the dotted lines), 4.5\,kpc (the solid lines), 6\,kpc (the dashed lines),
8\,kpc (the dot-dashed lines), and 10\,kpc (the double dot-dashed lines)}
\label{timedep}
\end{figure}

The evolution of the magnetic field energy, the cosmic ray energy and the mean
values of the magnetic pitch angles calculated in the subsequent rings for five
chosen distances from the galactic center is presented in Fig.~\ref{timedep}.

In all local models with nonvanishing differential rotation ( $q_{shear} \ne 0$)
i.e. at the radii $R>3\,kpc$, a fast growth  of the total magnetic field energy
is observed (see Fig.~\ref{timedep}, -- top panel). Within the ring
$R=4$--$5$\,kpc the mean growth time of the large-scale magnetic field, measured
in the phase of exponential amplification of the magnetic field (since
$t=0.2$\,Gyr till $t=0.9$\,Gyr) is the shortest -- about 0.15\,Gyr. The growth
time gradually increases with the radius up to about 0.3\,Gyr for the outermost
distance from the galactic center $R=10$\,kpc. The fastest growth of the
magnetic energy at 4.5\,kpc coincides with the largest SN rate and
high shear in comparison with other disk areas. For the smallest radii
($R<3$\,kpc), with assumed rigid rotation, the magnetic field is not amplified.
This justifies our decision of not taking into consideration rings close to the
center of the galaxy.


After the exponential amplification phase the magnetic field energy enters the
saturation phase -- the earlier, the quicker growth rate was observed. The
fastest growth of the CR energy (see Fig.\ref{timedep} -- middle panel) is
present at the radius 4.5\,kpc (solid line) due to the presence of the highest
SN rates there, similarly to the magnetic field energy behavior at this
galactocentric distance. In our simulations we do not obtain the energy
equipartition between the cosmic rays and the magnetic field. The CR energy
exceeds the magnetic energy by $1\div 2$ orders of magnitude. Possible reasons
for this excess are:  the low values of the CR diffusion coefficients,
as well as, periodic boundary conditions \cite[see][for more detailed
discussion]{hanasz08}. The bottom graph in Fig.~\ref{timedep} shows the time
evolution of the mean magnetic pitch angles at the same radii as for the
magnetic field and the cosmic ray energies. In the beginning stage of evolution
(until 0.2~Gyr) all curves exhibit maxima up to $10\degr$. Then, at two
distances from the galactic center, 4.5~kpc and 6~kpc, the values of the mean
magnetic pitch angles change sign to the negative ones. In the end, after
800~Myr, the pitch angles change again sign to positive values and stabilize for
all curves above $5^o$.

\begin{figure}
 \begin{center}
  \includegraphics[height=0.40\textheight]{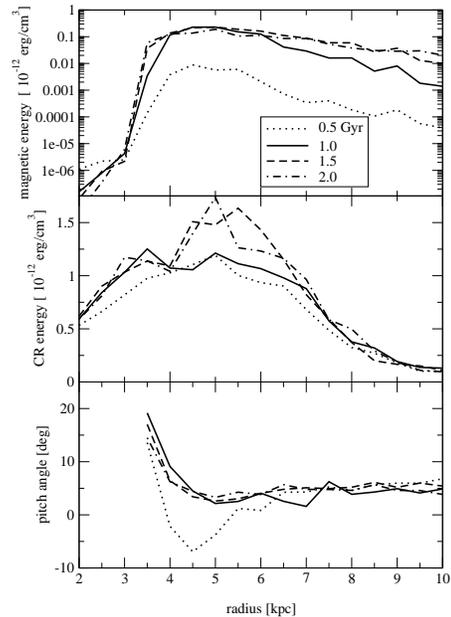}
 \end{center}
\caption{Radial dependency of the magnetic field energy, the cosmic ray energy
and the mean pitch angle values for four chosen time steps: 0.5\,Gyr (dotted
lines), 1\,Gyr (solid lines), 1.5\,Gyr (dashed lines),and  2\,Gyr (dot-dashed
lines).}
\label{radialdep}
\end{figure}

Fig.~\ref{radialdep} shows the radial dependency of the magnetic field energy
(top panel), the cosmic rays energy (middle panel), and the mean pitch angle
values (bottom panel) for four chosen time steps. We notice that the magnetic
field energy grows with time and decreases with the radius, as we present in the
Figure above. In addition, all curves (Fig~\ref{radialdep}, the middle panel)
showing the radial dependence of the cosmic ray energy exhibit the maxima in
their central parts (3--4\,kpc). The highest increase is observed for times
between 1.5 and 2\,Gyr and for the radius about 4.5~kpc. As we explained above,
the CR energy maxima are connected with the largest rate of SN between 4 and
5\,kpc (see Fig.~\ref{input}.

Due to small shear in the central part (see Fig.~\ref{input}), the mean magnetic
pitch angle (Fig.~\ref{radialdep}, bottom) reaches the highest values for all
chosen time steps. The smallest or negative pitch angle values are obtained at
time 0.5~Gyr at the radius 4.5~kpc (like in Fig.~\ref{timedep} -- bottom panel).
We can conclude that when the dynamo is still under development the pitch angles
are low or even negative. Later on, when the energy grows the pitch angles
attains about  $5^o$ for all radii.

\begin{figure}
 \begin{center}
  \includegraphics[height=0.40\textheight]{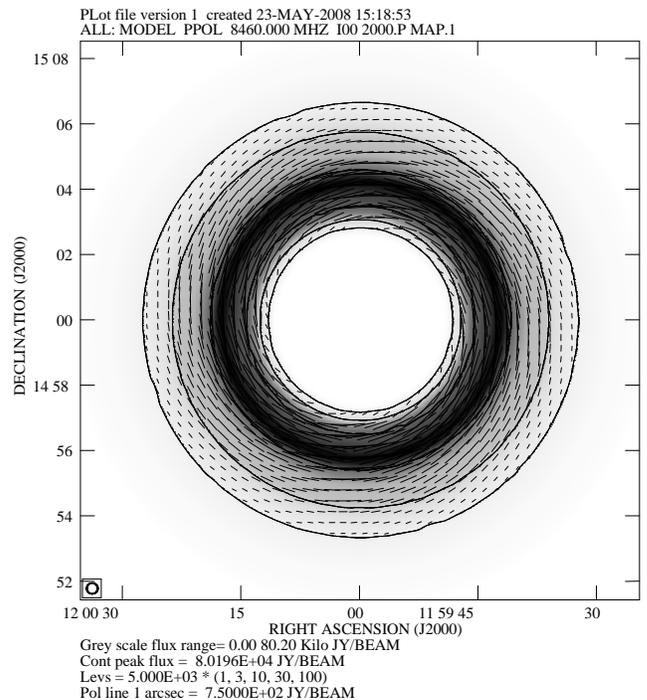}
 \end{center}
 \caption{Face-on map of the polarized emission of the modeled galaxy. Contours
and gray scale show polarized emission intensity, vectors are of directions of
apparent magnetic polarization vector, and length proportional to the intensity
}
\label{fig:face}
\end{figure}

The main goal of this paper is to compare the radio continuum emission obtained
from the cosmic ray dynamo model with the observed map of a real galaxy. In
Fig.~\ref{fig:face} we present the face-on view of our model map at the final
time of 2\,Gyr. The figure shows polarization vectors superimposed onto the
contours and grayplot of the polarized intensity. We see that the central part
exhibits the highest pitch angle values. Further out 
smaller pitch angles are observed, in agreement with the bottom pannels of 
Figs.~\ref{timedep} and \ref{radialdep}.
This fact can be explained byh the lower shear in the galactic center than
outward in the disk.

\begin{figure}
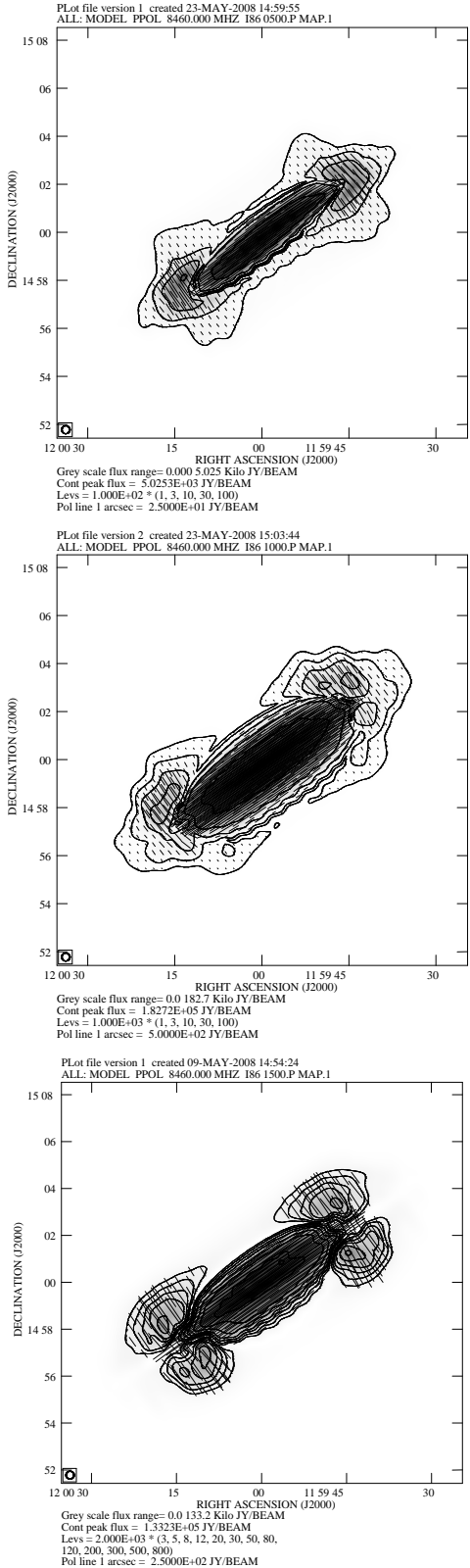

 \begin{center}
  \includegraphics[height=0.30\textheight]{f5a.eps}
  \includegraphics[height=0.30\textheight]{f5b.eps}
  \includegraphics[height=0.30\textheight]{f5c.eps}
 \end{center}
 \caption{Polarization maps of the modeled galaxy at selected time steps:
500\,Myr (top panel), 1000\,Myr (middle panel), and 1500\,Myr (bottom panel)
oriented as the real galaxy NGC~5775. Contours and gray scale show polarized
emission intensity, vectors are of directions of apparent magnetic polarization
vector, and length proportional to the intensity }
\label{fig:times}
\end{figure}

\begin{figure}
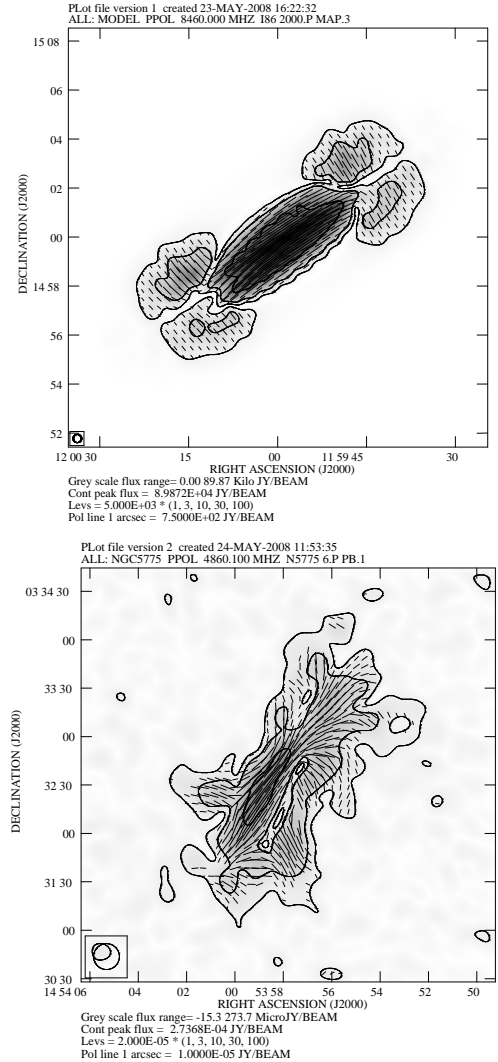

 \begin{center}
  \includegraphics[height=0.3\textheight]{f6a.eps}
  \includegraphics[height=0.3\textheight]{f6b.eps}
 \end{center}
 \caption{Map of the modeled galaxy at the latest (2\,Gyr) time-step of the
evolution (upper panel) oriented as the real galaxy NGC~5775 (lower panel).
Contours and gray scale show polarized emission intensity, vectors are of
directions of apparent magnetic polarization vector, and length proportional to
the intensity \label{fig:model}}
\end{figure}

Fig.~\ref{fig:times} shows calculated maps of the vectors of the polarized
emission superimposed onto the isolines and grayplots of the same quantity with
orientation on the sky-plane as the spiral galaxy NGC~5775 (inclination of
86$^\circ$ and position angle of 145$^\circ$) at selected time steps. One can
notice that the extended structures of the polarization vectors in the modeled
galaxy are present from the early stages (500~Myr) of the evolution. They appear
at large distances from the disk forming an X-shaped structure, similarly to the
observed map of the polarization emission of the edge-on galaxies. The most
extended structures are apparent at the late time steps of our simulations
1500~Myr (the bottom panel) and 2000\,Myr (see Fig.~\ref{fig:model}, the top panel).

In Fig.~\ref{fig:model} we compare maps of polarized emission for NGC~5775 (the
lower panel) and synthetic maps for our model of this galaxy (the upper panel).
In both pictures, isolines of polarized intensity are superimposed onto the
grayplot of the same quantity, together with the vectors of directions of the
apparent magnetic polarization with the length proportional to the intensity.
The polarization vectors in the central part of the bodies are parallel
to the disk in both modeled and real galaxies. The extended structures,
polarized perpendicularly to the disk plane, are visible in both maps as well.
However, in the real NGC~5775 the extensions are separated less than in our
model. The depolarized canals, clearly seen in the model map are barely visible
in eastern and western extensions in the real galaxy. This suggests that in the real
galaxy the magnetic field changes its orientation more smoothly than in the
model. Clear differences between the model and real observations appear close to
the major axis of the galaxy. Narrow depolarized channel is seen on the lower
panel of Fig.~\ref{fig:model}. It can be explained by relatively strong Faraday
effects caused by non-axisymmetric magnetic field configuration present in the
front side (the south-western side) close to the disk plane \citep{soida05}.
Another reason for this behavior could be connected to a local enhancement of
magnetic field and/or thermal electron density due to the influence of a spiral
arm. Our model does not include any non-axisymmetric features. Furthermore we do
not include the very central part in our modeled galaxy. This results in dumping
magnetic field at the radii $R<3$\,kpc. In real galaxies any non-axial
(vertical) motions can strongly affect magnetic field configuration.

\begin{figure}
 \begin{center}
  \includegraphics[width=0.4\textwidth]{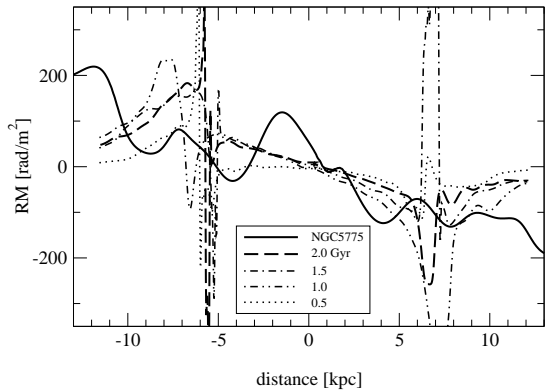}
 \end{center}
 \caption{Rotation measure plot along the major axis of the galaxy -- calculated
from the model at four time steps and the real observations of NGC~5775.
Negative distances denotes left (south-eastern) side of the galaxy }
\label{fig:rmplot}
\end{figure}

We compare our model with observations at relatively high frequency (4.86\,GHz),
where Faraday effects usually are small. Nevertheless we include such effects in
our calculation of Stokes parameters (Eq.~9). Calculating model maps of
polarized emission (intensity and directions) at two frequencies we were able to
construct the synthetic maps of rotation measures (RM) as well. In the RM
calculations we apply the density of gas taken directly from our simulations. A
cut of the rotation measure distributions along the major axis of both modeled
at four different time steps and observed galaxy (NGC~5775) for comparison are
presented in Fig.~\ref{fig:rmplot}. Our models reproduce the general trend of
the rotation measure. The deviations -- most prominent jumps around 6\,kpc of
both sides coincide with the narrow depolarized canals on the polarized
intensity map (Fig.~\ref{fig:model} -- upper panel), where the polarization
vectors change their direction rapidly (that leads to large uncertainties of RM
determination). Other differences between the model and observations (in the
center and at the peripheries) are possibly caused by non-axisymmetries  in
magnetic field and thermal gas distribution (such as spiral arms), certainly
present in NGC~5775 and not included in our model.

\section{Discussion}
\label{sec:discussion}

The extended structures appear in the modelled galaxy as a result of variety of
physical processes. First of all, the buoyancy driven by CRs transports the
magnetic field together with CR gas to the halo. We find that the winds in halo
build the vertical magnetic component. Those winds reach speeds up to
100\,km/s) and form quickly at the beginning stage of  \cite[see][]{hanasz04}.

Our local-box simulations show excessive energy density of CRs with respect to
to the magnetic energy. Possible reasons of this deviation from equipartition
are the reduced CR parallel and perpendicular diffusion coefficients and the
assumed horizontal periodicity of the computational domain. The latter
assumption leads to trapping of CRs by the horizontal magnetic field. 
\cite[see][]{hanasz08}.

Nevertheless, for the first time we apply the modeled CR distribution to perform
calculations of the polarized emission in our modeled galaxy. The obtained
results are in large extent in agreement with observations. The model presented
in this paper demonstrates that, even if the modeled disk starts from 2\,kpc
from the center of the galaxy, the magnetic field still can take the form of
X-shaped structures in the halo.

The maps of face-on model show that the pitch angles of the polarization vectors
are about 5$\degr$. This value is a bit smaller than the angles usually observed
in the spiral galaxies. We underline the fact, however, that our model does not
take into account effects of spiral arms in the galaxy. The arms understood as
manifestation of density waves are known to influence the magnetic field
structure in galactic disks significantly \citep{beck96}.

Our model calculates polarized intensity only. It does not include any turbulent
structures below relatively large grid size (20,\pc). This results in 
lack of total power emission map comaprable with real observations. It does not allow any reasonable discussion of depolarization other than caused
by beam smoothing and/or Faraday effects.

The Faraday effects computed in our model does not influence the overall
results. Differences between the rotation measure model and observations are
certainly due to limitations of axisymmetric model imposed by extending the
local-box into the global galactic model.

One could make the final picture more realistic via
the incorporation of supernova remnants. The incorporation of heat
output from supernovae may lower the cosmic ray excess. We suggest that global
simulations of cosmic ray-driven dynamo may solve the problem as well, however
both mentionned solutions of the CR excess problem require much larger
computational resources and other numerical algorithms than those currently
available for our project.
It seems however that the large-scale
qualitative features, namely the X-type structures can be succesfully reproduced
with the present setup. Therefore we would prefer to demonstrate the
observational properties of our present incomplete model of cosmic-ray driven
dynamo, rather than to mix quantitative solutions with  parametrized effects of
supernova remnants.
In the present simulations with the shearing box
approximation, it is not yet possible to model the vertical shear presumably
taking place in the case of NGC~5775 galaxy \citep{tul00}. Therefore the global
CR-MHD simulations of galactic disks free of limitations of the shearing box
should be addressed in the next step of development of our models.

Five galaxies NGC~891, NGC~3628, NGC~4217, NGC~4666, and NGC~5775 out of six
shown in \cite{soida05} characterized by the X-shaped structures of magnetic
field configuration, have flat rotation curve
\citep{sofue96,rhee96,mathew92,heald07}. One case (NGC~4631) exhibits different
magnetic field orientation with field lines crossing the disk plane. This galaxy
also is different when the rotation curve is taken into consideration. It
rotates rigidly to the large radii \citep{vauc63}. The lack of differential
rotation favors development of vertical structures at all galactic altitudes and
reduces the speed of transformation of the radial component of magnetic field
into the azimuthal one \cite{siejkowski08}. We suggest that the peculiar
structure of the polarized emission which is observed in NGC~4631, where vectors
cross the disk plane, can result from the cosmic-ray driven dynamo. This
statement is supported by a number of observations \cite{soida08}. The galaxy
NGC~4631 drew our attention and we plan to consider it in our future research.

\section{Conclusions}
\label{sec:conclusions}

In the present paper we compare the maps of polarized synchrotron radio emission
of galaxy NGC~5775 with those constructed from the numerical model of the cosmic
ray driven dynamo. The main conclusions can be summarized as follow:
\begin{itemize}
 \item All models of the local cosmic-ray driven dynamo computed at the galactic
radii between 4\,kpc and 10\,kpc indicate the fast growth of the magnetic flux
and the total magnetic energy.
 \item The synthetic radio maps of polarized emission computed on the basis of
our local models exhibit vertical magnetic field structures similar to those
observed in numerous edge-on galaxies.
 \item The polarization vectors in the disk plane (face-on) form a spiral
pattern with the pitch angles about 5$\degr$. The pitch angles from the model
are slightly smaller than we normally observe in galaxies. However, the pitch
angle of the mean magnetic field depends to some extent on the actual parameters
like e.g. the magnitude of the CR diffusion coefficients.
 \item As it was expected for observation at relatively high frequencies, the
Faraday effects considered in our model have no significant importance on the
final results.
 \item The major simplification of our model is caused by assumed axial
symmetry, which is necessary for construction of the global model from local-box
simulations.
\end{itemize}

\begin{acknowledgements}
This work was partly supported by the Polish Ministry of Science and Education
through the grants: 0656/P03D/2004/26 and 2693/H03/2006/31. Our collaboration
has been supported by Polish Ministry of Science and Education through the grant
ASTROSIM-PL.
\end{acknowledgements}

\end{document}